\documentclass[twocolumn,aps,amsmath,showpacs]{revtex4}
\usepackage[dvips]{graphics}
\newcommand{\be}{\begin{equation}}
\newcommand{\ee}{\end{equation}}

\newcommand{\bea}{\begin{eqnarray}}
\newcommand{\eea}{\end{eqnarray}}
\newcommand{\bd}{\begin{displaymath}}
\newcommand{\ed}{\end{displaymath}}
\newcommand{\bi}{\begin{itemize}}
\newcommand{\ei}{\end{itemize}}
\newcommand{\bc}{\begin{center}}
\newcommand{\ec}{\end{center}}
\newcommand{\bfl}{\begin{flushleft}}
\newcommand{\efl}{\end{flushleft}}
\newcommand{\bfr}{\begin{flushright}}
\newcommand{\efr}{\end{flushright}}
\newcommand{\f}{\frac}

\def\ra{\rightarrow}
\def\6{\partial}  
  \def\ve{\varepsilon}

\def\={\!\!\!&=&\!\!\!}
\def\+{\!\!\!&&\!\!\!+~}
\def\-{\!\!\!&&\!\!\!-~}

\begin{document}
\author{K. Brown$^{1}$, M. Crisan$^{2}$, and I. \c{T}ifrea$^{1}$}

\affiliation{$^1$Department of Physics and Astronomy, California State University, Fullerton, CA 92834, USA}
\affiliation{$^2$Department of Theoretical Physics, ``Babe\c{s}-Bolyai" University, 40084 Cluj-Napoca, Romania}

\date{\today}
\title{Transport and current noise characteristics of a T-shape double quantum dot system}


\begin{abstract}
We consider the transport and the noise characteristics for the case of a T-shape double quantum dot system using the equation of motion method. Our theoretical results, obtained in an approximation equivalent to the Hartree-Fock approximation, account for non-zero on-site Coulomb interaction in both the detector and side dots. The existence of a non-zero Coulomb interaction implies an additional two resonances in the detector's dot density of states and thereafter affects the electronic transport properties of the system. The system's conductance presents two Fano dips as function of the energy of the localized electronic level in the side dot. The Fano dips in the system's conductance can be observed both for strong (fast detector) and weak coupling (slow detector) between the detector dot and the external electrodes. Due to stronger electronic correlations the noise characteristics in the case of a slow detector are much higher. This setup may be of interest for the practical realization of qubit states in quantum dots systems.
\end{abstract}
\maketitle

\section{Introduction}

Transport through a complex quantum dot (QD) system is of high interest from both the practical applications and theoretical point of view \cite{tarucha}. From the applications point of view, QD systems may provide the perfect environment for the implementation of nanoelectronics and the realization of quantum bits (qubits). On the other hand, QD systems allow the theoretical study of quantum many body effects. Particulary, the Anderson single impurity model \cite{anderson} was extensively used to understand the electronic correlations in QD systems. The model, in which QD's are represented as impurities, was successfully applied to the study of single or multiple-dot systems. For example, single QD's systems allow the controlled realization of the Kondo regime of the Anderson impurity problem \cite{cronenwett}. Multiple-dot systems may be subject to inter-dot coupling, and accordingly novel many body states can be generated.

In the case of a double-QD system, with the QD's arranged in a series, parallel, or T-shape configuration (see Ref. \cite{kawakami} for a picture of these configurations), it was shown that the arrangement of the component QD's plays an important role when transport properties are investigated \cite{kawakami}. The results obtained for the system's conductance can be explained based on Kondo resonances influenced by Fano interference effects \cite{fano}. In the case of a series configuration there are no interference effects and the possibility of a double peaked Kondo resonance due to the inter-dot tunneling may reduce the system's conductance. Differently, in the case of a parallel configuration the two different channels of electron propagation are responsible for a sharp and a broad Kondo resonance. The interference between these two Kondo resonances will significantly reduce the system's conductance \cite{kawakami}. The double dot T-shape configuration, with one dot (detector dot) directly connected to the external leads and the second dot (side dot) coupled to the first one but not to the external leads, has a particular transport behavior. In this case, the density of states (DOS) of the detector dot (coupled to the external leads) has a broad resonance and develops a sharp dip structure due to the interference with the states from the second dot while the second dot DOS presents a sharp Kondo resonance. Accordingly, the system's conductance is very small in the Kondo regime. A similar double dot system, with one Kondo dot and one effectively noninteracting dot, was shown by Dias da Silva {\em et al}. \cite{dias} that it can be continuously tuned to create a pseudogapped DOS and access a quantum-critical point separating Kondo and non-Kondo phases. Different approaches for the study of the T-shape system were based on the Anderson impurity model. For example, Wu {\em et al}. \cite{wu} considered an infinite on-site Coulomb interaction in the detector dot  so the double occupancy was forbidden in this dot. On the other hand, Guclu {\em et al}. \cite{guclu} assumed an infinite on-site Coulomb interaction in the side QD, neglecting it in the detector QD. Although the two mentioned configurations are different, both the Kondo effect and the Fano interference effect play an important role in the system's transport properties \cite{wu,guclu}. Transport properties in quantum dot systems were discussed also in terms of anti-resonance scattering, a method which gives similar results with the standard Green's function method \cite{wang}. In short, the evaluation of the system's transmission coefficient shows that for a single impurity Anderson model every time the conduction electron's energy is equal to the localized level energy, an anti-resonance scattering occurs, leading to a dip in the system's conductance. In this case the transmission coefficient vanishes and the conduction channel is completely blocked by the presence of the localized (impurity) level.

\begin{figure}[tb]
\centering \scalebox{0.35}[0.35]{\includegraphics*{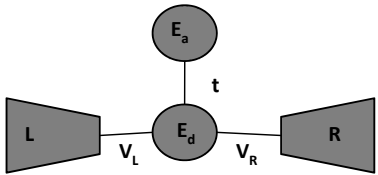}}
\caption{Schematic representation of the T-shape double quantum dot system. The detector dot (characteristic energy $E_d$) is coupled both to the side quantum dot (characteristic energy $E_a$) and the external electrodes $L$ and $R$.}
\label{fig0}
\end{figure}

Here we propose an investigation of the T-shape double quantum dot system (See Fig. \ref{fig0}) based on the equation of motion (EOM) method. Our approximation is similar to the one used by Hewson \cite{hewson} in the study of the Anderson's single impurity model. We will consider the general case with finite on-site Coulomb interaction in both the detector and side dots. We will discuss the main transport properties of the system including the system's conductance and current noise characteristics. The paper is organized as follows: In Section II we present the general Hamiltonian of the system and using the EOM method we estimate the detector's dot DOS. In Section III we analyze the system's main electronic transport properties. Finally, Section IV presents our conclusions.

\section{Model}

The T-shape double quantum dot system is described by the following general Hamiltonian:
\bea
H&=&\sum_{k,\sigma;\alpha}\ve_k c^\dagger_{k\sigma;\alpha}c_{k\sigma;\alpha}+
\sum_\sigma E_d d^\dagger_\sigma d_\sigma+U_d n_{d\uparrow}n_{d\downarrow} \nonumber\\
&+& \sum_\sigma E_a a_\sigma^\dagger a_\sigma + U_a n_{a\uparrow}n_{a\downarrow} +
t\sum_\sigma\left(d_\sigma^\dagger a_\sigma+a^\dagger_\sigma d_\sigma\right)\nonumber\\
&+&\sum_{k,\sigma;\alpha} V_{kd;\alpha}\left(c^\dagger_{k\sigma;\alpha}d_\sigma+d^\dagger_\sigma c_{k\sigma;\alpha}\right)\;.
\eea
The first term in the Hamiltonian describes the free electrons in the leads, $c^\dagger_{k\sigma;\alpha}$ and $c_{k\sigma;\alpha}$ being fermionic creation and annihilation operators for electrons with momentum $k$ and spin $\sigma$ in the lead $\alpha$ ($\alpha\equiv$ left (L), right (R)). The following four terms describe the mesoscopic part of the Hamiltonian and correspond to the electrons localized in the detector ($E_d$) and side ($E_a$) quantum dots of the system. Additionally, electrons in each component dot are subject to on-site Coulomb interaction described by the interaction terms $U_d$ and $U_a$, respectively. The last two terms in the Hamiltonian describe interactions between the system's electrons. The coupling constant $t$ characterizes the interaction between the electrons localized in the detector and side dots of the system. $V_{kd;\alpha}$ characterizes the interaction between the free electrons in the lead $\alpha$ and the localized electrons in the detector dot. For simplicity we will consider the case $V_{kd;L}=V_{kd;R}$ in which the detector couples to the leads only in the symmetric combination $c_{k\sigma}=(c_{k\sigma;L}+c_{k\sigma;R})/\sqrt{2}$ and the dot connects effectively to a single lead, with $V_{kd}=\sqrt{2}V_{kdL}$.

The system's transport properties can be investigated using the Green's function formalism. The main quantity will be the Green's function corresponding to the localized electrons in the detector dot. One way to extract this Green's function is to use the EOM method.  It is well known that in the case of a general Anderson impurity model, the EOM method leads to an infinite hierarchy of higher-order Green's functions, so in order to obtain the detector's dot electronic Green's function  one needs to introduce a reliable approximation to truncate this hierarchy. The difficulty is mainly introduced by the interaction terms in the system's Hamiltonian. When the on-site Coulomb interaction term is absent, an exact solution of the problem is possible as it is well known that in this case the set of equations obtained from the EOM method are closed. In the case of two fermionic operators $A$ and $B$ the Fourier transform of the Green's function with respect to the time, $G_{AB}(\omega)=\left<\left<A;B\right>\right>$, is given by the general equation
\be
\omega\left<\left<A;B\right>\right>=\left<\left\{A,B\right\}\right>+\left<\left<\left[A,H\right];B\right>\right>\;,
\ee
where $\left<A\right>$ represents the mean value of the operator $A$, $\left\{A,B\right\}$ the anti-commutator of the operators $A$ and $B$, and $\left[A,B\right]$ their commutator. The last term on the right hand side of the equation is responsible for the generation of the infinite chain of higher order Green's functions. Our main goal will be to calculate the $d$-electrons Green's function, $G^\sigma_dd(\omega)=\left<\left<d_\sigma;d^\dagger_\sigma\right>\right>$, to be used in the estimation of the system's transport properties. The calculations are relatively simple, however, they lead to a chain of coupled Green's functions equations. The approximation we used to close this set of equations is similar to the one introduced by Hewson \cite{hewson} and in some sense is equivalent to the standard Hartree-Fock approximation. The resulting Green's functions for the localized electrons in the detector and side dots are given by
\begin{widetext}
\be\label{gdd}
G^\sigma_{dd}(\omega)=\left[\f{(\omega-E_d)(\omega-E_d-U_d)}{\omega-E_d-(1-<n_{d-\sigma}>)U_d}-
\sum_k\f{|V_{kd}|^2}{\omega-\ve_k}-t^2\f{\omega-E_a-(1-<n_{a-\sigma}>)U_a}{(\omega-E_a)(\omega-E_a-U_a)}
\right]^{-1}
\ee
and
\be\label{gaa}
G^\sigma_{aa}(\omega)=\left[\f{(\omega-E_a)(\omega-E_a-U_a)}{\omega-E_a-(1-<n_{a-\sigma}>)U_a}-
\f{t^2}{\f{(\omega-E_d)(\omega-E_d-U_d)}{\omega-E_d-(1-<n_{d-\sigma}>)U_d}-
\sum_k\f{|V_{kd}|^2}{\omega-\ve_k}}
\right]^{-1}\;,
\ee
\end{widetext}
where $<n_{d-\sigma}>$ and $<n_{a-\sigma}>$ represent the average occupancy of the electronic levels in the two dots of the system. Note that in the case $t=0$ the general case of a single Anderson impurity is recovered. The above equations are coupled as based on the general Green's function formalism the average occupancy of the electronic level is given by
\be
\left<n_{d\sigma}\right>=-\f{1}{\pi}\int \textrm{Im}G^{\sigma}_{dd}(\omega+i\eta)d\omega\;,
\ee
with $\eta\ra 0$. A similar relation stands for $<n_{a\sigma}>$. The imaginary part of the Green's functions can be extracted if we use the general relation
\begin{figure}[b]
\centering \scalebox{0.75}[1]{\includegraphics*{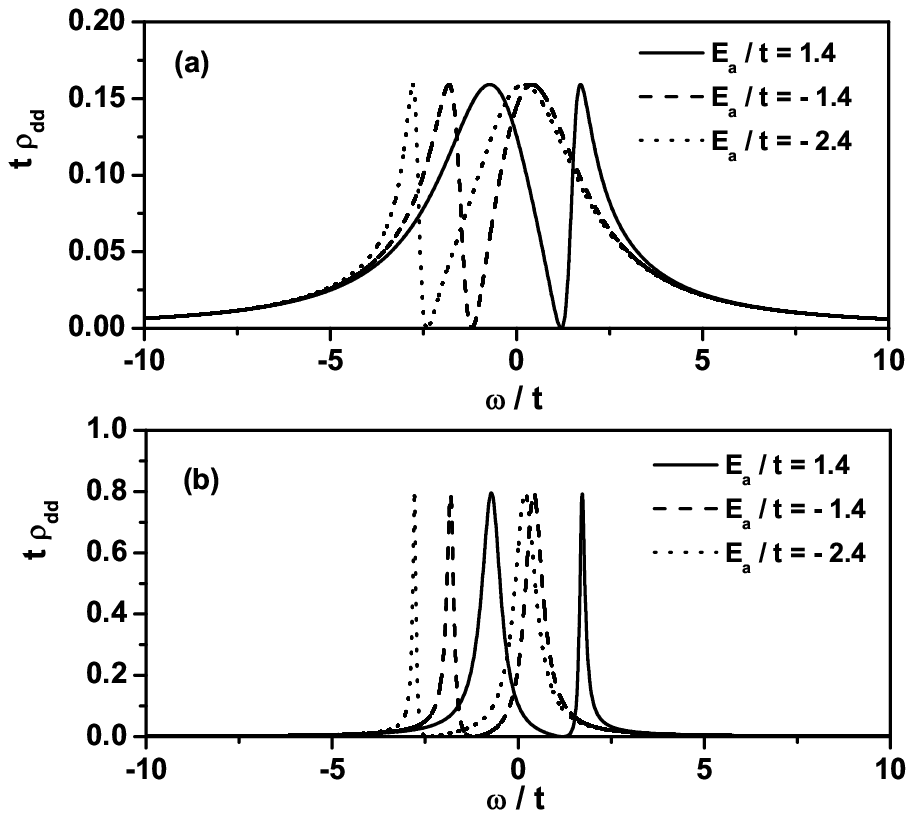}}
\caption{The detector's dot DOS in the absence of on-site Coulomb interaction ($U_d=U_a=0$) for the case of a (a) fast detector ($\Delta/t=2$) and a (b) slow detector ($\Delta/t=0.4$) for different values of the ratio $E_a/t$ ($E_a/t=1.4$ - full line, $E_a/t=-1.4$ - dashed line, and $E_a/t=-2.4$ - dotted line). For both cases $E_d/t=-0.2$.}
\label{fig1}
\end{figure}
\bd
\sum_k\f{|V_{kd}|^2}{\omega-\ve_k+i\delta}={\cal{P}}\sum_k\f{|V_{kd}|^2}{\omega-\ve_k}-i\Delta\;,
\ed
where $\cal{P}$ represents the principal part and we introduced for convenience the notation $\Delta=\pi\sum_k |V_{kd}|^2\delta(\omega-\ve_k)$ with $\delta(x)$ the delta-Dirac function. By definition the detector dot DOS is given by
\begin{equation}\label{defDOS}
\rho_{dd}^\sigma(\omega)=-\f{1}{\pi}\textrm{Im}G^\sigma_{dd}(\omega)\;.
\end{equation}
The simplest situation occurs when the on-site Coulomb interaction in both the detector and side dots is neglected $U_d=U_a=0$. In this case, the Green's function are not coupled and the detector dot density of state becomes
\begin{equation}
\rho^\sigma_{dd}(\omega)=\f{1}{\pi}\f{\Delta}{\left(\omega-E_d-\f{t^2}{\omega-E_a}\right)^2+\Delta^2}\;.
\end{equation}
\begin{figure}[b]
\centering \scalebox{0.75}[1]{\includegraphics*{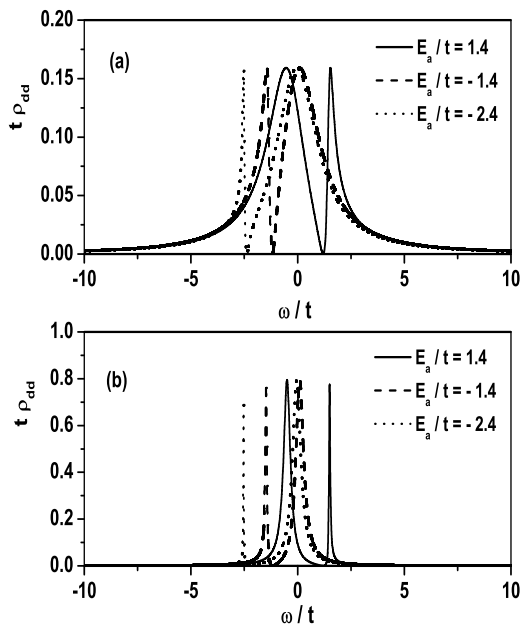}}
\caption{The detector's dot DOS in the presence of infinite on-site Coulomb interaction ($U_d=U_a=\infty$) for the case of a (a) fast detector ($\Delta/t=2$) and a (b) slow detector ($\Delta/t=0.4$) for different values of the ratio $E_a/t$ ($E_a/t=1.4$ - full line, $E_a/t=-1.4$ - dashed line, and $E_a/t=-2.4$ - dotted line). For both cases $E_d/t=-0.2$.}
\label{fig2}
\end{figure}
It is convenient to introduce dimensionless quantities and measure all energies in units of the coupling constant $t$ relative to the Fermi level of the conduction electrons in the leads ($E_F=0$). We will analyze two different situations relative to the transport properties of the T-shape double quantum dot system. In the first case $\Delta/t>1$ and the conduction electrons will flow relatively fast through the detector dot and their interactions with the localized electrons in the side dot will not influence drastically the transport properties. On the opposite situation, when $\Delta/t<1$, the flow of the conduction electrons will depend on the interaction with the localized electrons in the side dot and therefore we expect this situation to reflect on the transport properties of the system. We will address the case of a large ratio ($\Delta/t=2$) as a fast detector and the case of a small ratio ($\Delta/t=0.4$) as a slow detector.

In Figure \ref{fig1} we present the DOS for the detector dot for both the fast and slow detector configurations. As a general result, the ratio $\Delta/t$ determines the full width half maximum (FWHM) value of the density of states. One can clearly see that the density of states presents sharper peaks for the case of a slow detector. The value of the energy level in the side dot ($E_a$) controls the position of the two peaks in the detector's dot DOS. Our calculation is performed at $T=0$ K , however, we expect that effects related to temperature to be minimal on the DOS function. For both cases $E_d/t=-0.2$.

\begin{figure}[t]
\centering \scalebox{0.8}[0.9]{\includegraphics*{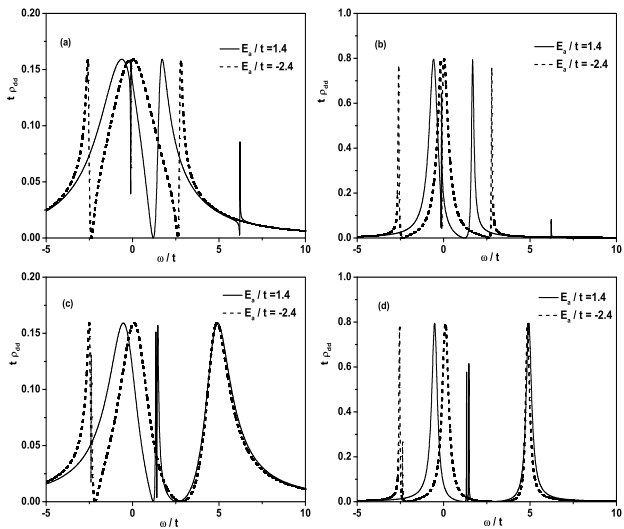}}
\caption{The detector's dot DOS for finite on-site Coulomb interaction ((a) and (b) -- $U_d/t=0.2$, $U_a/t=5$; (c) and (d) -- $U_d/t=5$, $U_a/t=0.2$) for the case of a fast detector ((a) and (c) -- $\Delta/t=2$) and for a slow detector ((b) and (d) -- $\Delta/t=0.4$) for different values of the ratio $E_a/t$ ($E_a/t=1.4$ - full line, $E_a/t=-2.4$ - dashed line). For both cases $E_d/t=-0.2$.}
\label{fig3}
\end{figure}

A similar situation occurs for infinite on-site Coulomb interaction ($U_d=U_a=\infty$). In this case, due to strong Coulomb repulsion between electrons, only one electron can occupy any of the two localized energy levels in the detector and side dots. The two equations giving the electron Green's functions in the system's component dots are coupled via the terms involving the occupation numbers $<n^\sigma_d>$ and $<n^\sigma_a>$. Although a solution for the two Green's functions requires a self-consistent calculation, the advantage of the infinite on-site Coulomb interaction case is that the limit $U_d=U_a=\infty$ the form of the two equations is simplified. To solve these equations, we start from the Green's functions corresponding to the $U_d=U_a=0$ case and use them as a starting point to calculate the occupancy of the side dot level $<n^\sigma_a>$. Thereafter, we use this value to start the self-consistent calculation of the two Green's functions using Eqs. (\ref{gdd}) and (\ref{gaa}). The results of the calculation are presented in Figure \ref{fig2}. In the case of infinite on-site Coulomb interaction the detector dot DOS has the same number of peaks as in the previous case, however, in this situation the peaks are grouped together.

A more interesting case is the one with arbitrary on-site Coulomb interaction in the two component dots. A non-zero on-site Coulomb interaction is responsible for two additional peaks in the detector dot DOS. The position of these peaks is determined by the relative strength of the on-site Coulomb interaction. Figures  \ref{fig3}a and \ref{fig3}b considers the detector dot DOS for $U_d/t=0.2$ and $U_a/t=5$ and Figures \ref{fig3}c and \ref{fig3}d for $U_d/t=0.2$ and $U_a/t=5$. Each case is analyzed for $E_d/t=-0.2$ and  $E_a/t=1.4$ (full line) and $E_a/t=-2.4$ (dashed line). The presence of additional peaks in the detector's dot density of states will reflect in the transport properties of the system.

\section{Transport properties}

Transport properties for the T-shape double quantum dot system can be discussed in terms of system's conductance and current noise characteristics. To calculate the system's conductance we use the Meir and Weingreen \cite{meir} formula:
\begin{equation}
G=G_{0}\int_{-\infty}^{\infty} d\omega\;\frac{\Delta}{2} \left[-\frac{\partial f(\omega)}{\partial\omega}\right]\rho_{dd}(\omega)\;,
\end{equation}
where $G_{0}=2e^{2}/\hbar$ ($e$ is the electron charge and $\hbar$ the Plank constant), $\rho_{dd}(\omega)$ is given by Eq.(\ref{defDOS}), and $f(x)$ represent the Fermi-Dirac distribution function. This general equation will lead to an expression for the system's conductance as function of temperature. We will limit our study to the $T=0$ K case, when the derivative of the Fermi-Dirac function can be substituted by a Dirac-delta function. In the case $t=0$, i.e., when the side dot is decoupled from the system, the general transport theory for the Anderson single impurity model predicts the existence of a Kondo peak in the system's conductance. The situation completely changes for $t\neq 0$.

\begin{figure}[t]
\centering \scalebox{0.8}[0.9]{\includegraphics*{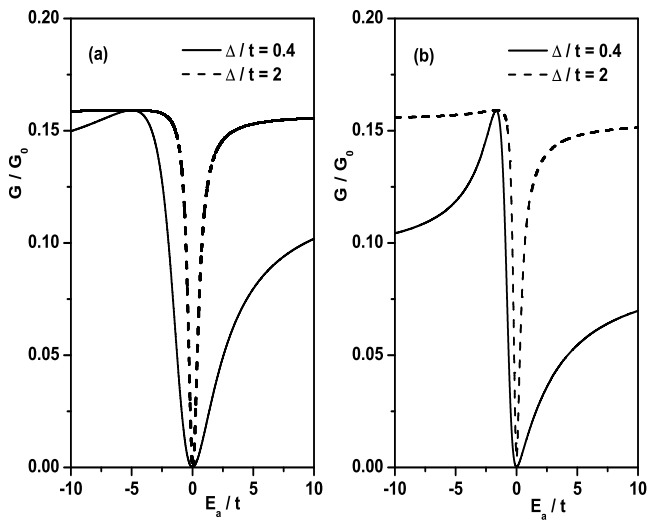}}
\caption{The system's conductance as function of the energy level in the side dot for (a) $U_d=U_a=0$ and (b) $U_d=U_a=\infty$ in the case of a slow detector $\Delta/t=0.4$ (full line) and in the case of a fast detector $\Delta/t=2$ (dashed line). For both cases $E_d/t=-0.2$.}
\label{fig4}
\end{figure}

The trivial case without on-site Coulomb interaction ($U_d=U_a=0$) can be solved analytically. In this case the localized electron's Green's function can be obtained exactly and the system's conductance is given by
\be
G=G_0\f{\Delta^2}{2\pi}\f{E_a^2}{\left(E_dE_a-t^2\right)^2+\Delta^2E_a^2}\;.
\ee
When the two dots are decoupled from the leads ($\Delta=0$) the system's conductance vanishes and when the side dot is decoupled from the detector dot ($t=0$) the system's conductance matches the conductance obtained from the Anderson single impurity model \cite{cornaglia}. The situation is slightly different when we consider a nonzero on-site Coulomb interaction in both the detector and side dots of the system. In this case, an exact analytical solution is impossible, approximations being required to obtained the localized electrons Green's function. To account for the system's transport properties we used the approximation introduced by Hewson \cite{hewson} which is very similar to the Hartree-Fock approximation. The disadvantage of this approximation is that it cannot account for the physics related to the Kondo effect. Figure \ref{fig4} presents the system's conductance as function of the side dot energy level for a) $U_d=U_a=0$ and b) $U_d=U_a=\infty$. In both situations when the side dot energy $E_a$ is closed to the Fermi level ($E_F=0$) an anti-resonance scattering occurs leading to a sharp drop in the system's conductance. As a general result, the dip is sharper in the case of infinite on-site Coulomb interaction and a fast detector configuration.

\begin{figure}[t]
\centering \scalebox{0.8}[0.9]{\includegraphics*{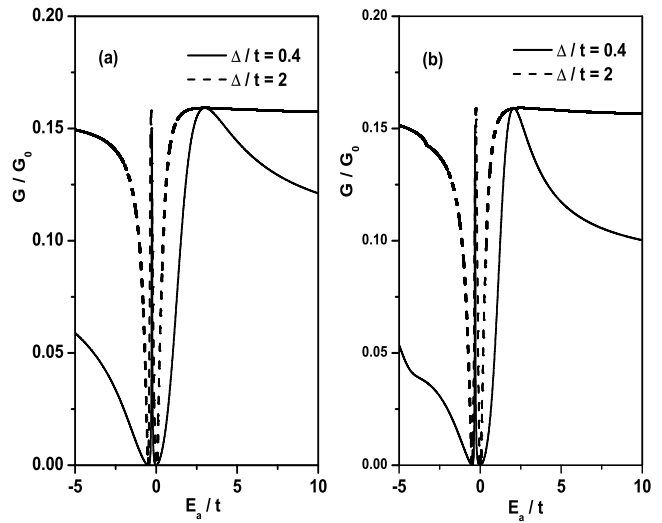}}
\caption{The system's conductance as function of the energy level in the side dot for (a) $U_d/t=5$ and $U_a/t=0.2$ and (b) $U_d/t=0.2$ and $U_a/t=5$ in the case of a slow detector $\Delta/t=0.4$ (full line) and in the case of a fast detector $\Delta/t=2$ (dashed line). For both cases $E_d/t=-0.2$.}
\label{fig5}
\end{figure}

A completely different situation occurs when finite on-site Coulomb interaction is considered in both the detector and side dots. In this case, the system's conductance presents a double dip structure corresponding to two possible anti-scattering processes corresponding to each transport channel. The presence of the two dips is consistent and does not depend on the relative size of the on-site Coulomb interaction in the two component dots. Figure \ref{fig5} presents the conduction of the double dot T-shape system for different values of the side dot energy level. We consider two different situation, $U_d/t=5$ and $U_a/t=0.2$ for Figure \ref{fig5}a and $U_d/t=0.2$ and $U_a/t=5$ for Figure \ref{fig5}. In both cases we consider the ratio $E_d/t=-0.2$, however, different values for this ratio do not change qualitatively the behavior of the system's conductance.

Quantum and thermal fluctuations are very important as they are one of the reasons why quantum correlations are difficult to observe experimentally. Such fluctuations can be estimated based on the current noise characteristics.
We will discuss the system's current noise characteristics in terms of the Fano factor defined as the ratio of the shot noise $S(V)$ and the current $I(V)$ passing through the system, i.e., $\gamma=S(V)/(2eI(V))$, $e$ being the electron charge. The current shot noise, $S(V)$, when an external bias $V$ is applied to the detector dot is defined  as a correlation function of current fluctuations and it can be proved that is related to the detector dot DOS via the transmission function, $T(\omega)=\pi\Delta\rho_d(\omega)$ \cite{lopez}:
\begin{equation}
S(V)=\frac{4e^2}{h}\int_{-eV/2}^{eV/2} d\omega\; T(\omega)\left[1-T(\omega)\right]\;,
\end{equation}
where $h$ is the Planck's constant. On the other hand, the source current can be calculated as:
\begin{equation}
I_L=\f{2e}{h}\int_{-\infty}^\infty d\omega\;T(\omega)\left[f_L(\omega)-f_R(\omega)\right]\;,
\end{equation}
where $f_\alpha(\omega)=\left[\exp{\{(\omega-\mu_\alpha)/k_B T\}}+1\right]^{-1}$ with $T$ being the temperature and $k_B$ the Boltzmann constant. For a symmetrical bias condition we set $\mu_L=E_F-eV/2$ and $\mu_R=E_F+eV/2$ with $E_F=0$. Note that in our particular situation the current is conserved, $I_L=I_R$. In the $T=0$ K limit the Fermi-Dirac functions $f_\alpha(\omega)$ are given by usual step functions and the evaluation of the current is relatively simple. Special attention should be given to the point $V=0$ when the Fano factor can be calculated as $\gamma=1-T(E_F)$.

\begin{figure}[t]
\centering \scalebox{0.8}[0.9]{\includegraphics*{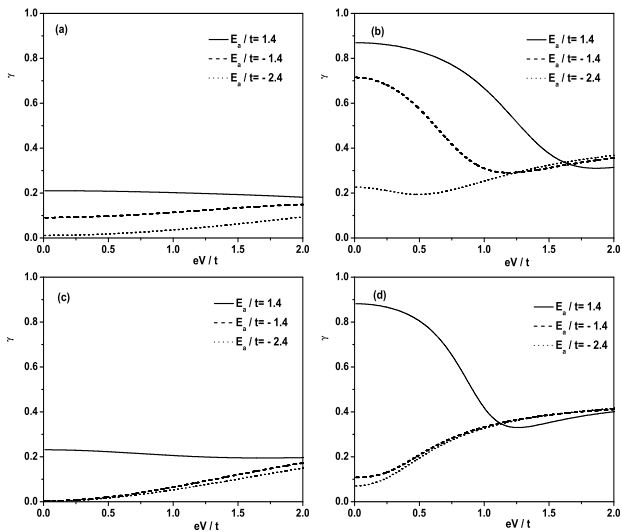}}
\caption{The system's Fano factor $\gamma$ as function of the applied bias for (a)-(b) $U_d=U_a=0$ and (c)-(d) $U_d=U_a=\infty$ in the case of a fast detector $\Delta/t=0.4$ ((a) and (c)) and in the case of a slow detector  $\Delta/t=2$ ((b) and (d)) for various values of the ration $E_a/t$ ($E_a/t=1.4$ - full line, $E_a/t=-1.4$ - dashed line, and $E_a/t=-2.4$ - dotted line). For all situations $E_d/t=-0.2$.}
\label{fig6}
\end{figure}
\begin{figure}[t]
\centering \scalebox{0.8}[0.9]{\includegraphics*{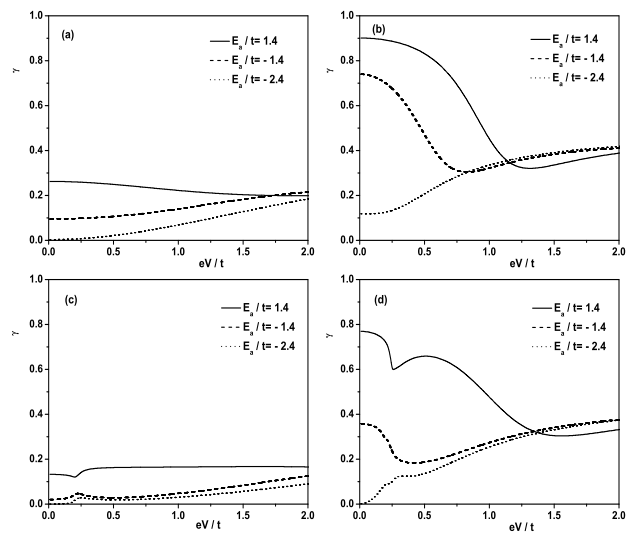}}
\caption{The system's Fano factor $\gamma$ as function of the applied bias for (a)-(b) $U_d/t=5$ and $U_a/t=0.2$ and (c)-(d) $U_d/t=0.2$ and $U_a/t=5$ in the case of a fast detector $\Delta/t=0.4$ ((a) and (c)) and in the case of a slow detector  $\Delta/t=2$ ((b) and (d)) for various values of the ration $E_a/t$ ($E_a/t=1.4$ - full line, $E_a/t=-1.4$ - dashed line, and $E_a/t=-2.4$ - dotted line). For all situations $E_d/t=-0.2$.}
\label{fig7}
\end{figure}

In Figure \ref{fig6} we plotted the Fano factor as function of the applied external bias for (a)-(b) $U_d=U_a=0$ and (c)-(d) $U_d=U_a=\infty$. We considered both the slow and fast detector situations for a fixed energy level in the detector dot ($E_d/t=-0.2$) and different values for the energy level in the side dot ($E_a/t=1.4$ - full line, $E_a/t=-1.4$ - dashed line, and $E_a/t=-2.4$ - dotted line). As a general result, the Fano factor for the case of a slow detector is relatively larger than that of a fast detector. The reason for this behavior is due to the difference in the current magnitude between the slow and fast detector configurations, as for both cases the shot noise has the same order of magnitude. A similar situation occurs also when we consider finite on site interaction in both the detector and side dots (see Fig \ref{fig7}). From the microscopic point of view, the larger Fano factor in the slow detector is explained by the stronger coupling of the conduction electrons to the energy level in the side dot. In all situations the Fano factor is relatively larger when the on-site Coulomb interaction is non-zero, indicating that stronger electronic interactions in the system are responsible for more noise.

\section{Conclusions}

In conclusion, we presented a theoretical investigation of the main transport properties of a T-shape double dot system based on the EOM method. Our results are obtained in an approximation equivalent to the Hartree-Fock approximation for zero, finite, and infinite on-site Coulomb interaction in both the detector and side dots. We considered two possible configurations of the system based on the relative interaction strength between the conduction electrons and the localized electrons in the detector dot. In the slow detector situation, this interaction is smaller than the interaction between the electrons in the detector and side dots ($\Delta/t<1$) and we proved that the transport properties of the system are strongly affected by the presence of the side dot. On the other hand, in the case of a fast detector, when the interaction between the electrons in the detector and side dots is relatively small compared to the interaction between the conduction electrons and the electrons localized in the detector dot ($\Delta/t>1$), the role of the side dot is diminished in the system's transport properties. The first quantity we considered was the detector dot DOS. When the on-site Coulomb interaction is set to zero or infinite in both dots, the detector dot DOS shows two peaks, slightly sharper for the case of a slow detector. The situation is different when the on-site Coulomb interaction has finite value in both the component dots. In this case, additional peaks are present in the detector dot DOS. Still, in the case of a slow detector, these peaks are sharper. The second quantity we considered was the system's conductance. As a general result, the system's conductance presents an oscillatory behavior as function of the energy level in the system's side dot. For the zero and infinite on-site Coulomb interaction, there is only one oscillation represented by a dip in the system's conductance. When the on-site Coulomb interaction is finite in both dots, the system's conductance presents two dips as a signature for the additional peaks in the detector's dot DOS. As a general result, the dips in the system's conductance are sharper for the case of a fast detector. This situation can be understood if we consider the derivative of the Fermi function in the system's conductance which can account  for all sharp features in the detector's dot DOS and lead to a broader conductance for the case of a slow detector. Finally, we analyzed the noise in the system's transport in terms of the Fano factor. Our results prove that when electronic correlations are stronger the noise is higher, making the signal detection in the slow detector case more difficult to measure. The Hartree-Fock approximation leads to results in good agreement with the results generated by the bosonisation method, at least when we discuss the system's main transport properties. On the other hand, an approximation which includes terms beyond the Hartree-Fock approximation has to be considered in connection with other possible phenomena in quantum dot systems. One example is the Kondo effect whose physics can be investigated only if we consider additional higher order Green's functions in the equation of motion method \cite{lacroix}. Another example is the investigation of the differential capacitance in quantum dots or molecular systems \cite{wang2,fan}. A higher order approximation is required for the understanding of the system's transport properties when a transition between the Coulomb blockade regime to the Kondo regime is considered \cite{natalya}. Especially, it will be of great interest to account for such effects in multi-quantum dots systems \cite{tifrea}.

\begin{acknowledgments}
MC would like to acknowledge financial support from the Romanian National Research Program PN II-ID-502. Two of the authors, M.C. and I.T., would like to thank A. Aldea for helpful discussions.
\end{acknowledgments}

\end{document}